\documentclass[fleqn,usenatbib]{mnras}

\usepackage[T1]{fontenc}
\usepackage{ae,aecompl}

\usepackage{graphicx}	
\usepackage{amsmath}	
\usepackage{amssymb}	

\newcommand{\msun}{\hbox{$M_{\odot}$}}

\newcommand{\re}{\hbox{${\rm R}_{\rm e}$}}

\title[Quenching and structure of red nuggets]{Star formation quenching imprinted on the internal structure of naked red nuggets}

\author[I. Mart\'in-Navarro]{
Ignacio Mart\'in-Navarro$^{1,2}$\thanks{E-mail: imartin@mpia.de}, Glenn van de Ven$^{3,4}$ and Ak{\i}n Y{\i}ld{\i}r{\i}m$^{5}$
\\
$^{1}$University of California Santa Cruz, 1156 High Street, Santa Cruz, CA 95064, USA\\
$^{2}$Max-Planck Institut f\"ur Astronomie, Konigstuhl 17, D-69117 Heidelberg, Germany\\
$^{3}$European Southern Observatory, Karl-Schwarzschild-Str. 2, 85748 Garching b. M\"unchen, Germany\\
$^{4}$University of Vienna, Department of Astrophysics, T\"urkenschanzstrasse 17, 1180 Vienna, Austria\\
$^{5}$Max-Planck Institut f\"{u}r Astrophysics, Karl-Schwarzschild-Str. 1, 85741 Garching, Germany\\
}

\date{Accepted XXX. Received YYY; in original form ZZZ}

\pubyear{2018}

\begin{document}
\label{firstpage}
\pagerange{\pageref{firstpage}--\pageref{lastpage}}
\maketitle

\begin{abstract}

The formation and assembly process of massive galaxies is a combination of two phases: an initial {\it in-situ}-dominated one followed by an {\it ex-situ}-dominated evolution. Separating these two contributions is therefore crucial to understand the baryonic cycle within massive halos. A recently discovered population of so-called naked red nuggets, galaxies that shortcut the {\it ex-situ} stage preserving their pristine properties, presents a unique opportunity to study in detail star formation in massive galaxies without the confounding effect of later accretion. We investigate the spatially resolved star formation histories of a sample of 12 naked red nuggets. We measure how their radial light distributions, star formation rates and central densities evolved in time. We find that, while forming stars, red nuggets become gradually more concentrated, reaching a maximum concentration at quenching. After being quenched, they kept forming stars in a more disky-like configuration. Our measurements suggest that super-massive black holes and host galaxies grow their mass in a self-regulated way until a characteristic M$_\bullet$/M$_\mathrm{halo}$ is reached. Once black holes are massive enough, red nuggets get quenched and depart from the star formation main sequence. While in the main sequence, red nuggets evolve at roughly constant star formation rate. This can explain up to $\sim0.3$ dex of the scatter of the star formation main sequence, as well as its higher normalization observed in the early Universe. Hence, our results suggest that the main sequence is composed of populations of galaxies at different evolutionary stages, and that the scatter is therefore due to secular processes.

\end{abstract}



\section{Introduction}

\vspace{0.04in}

In the local Universe, the high-mass end of the galaxy mass function consists almost entirely of early-type galaxies (ETGs) \citep[e.g.][]{Kochanek01,Bell03,Kauffmann03,Blanton05,Baldry04,Baldry06,Vulcani11,Kelvin14}. ETGs are characterized by two main properties, their structure and their stellar populations, which tightly constrain the formation and evolution channels of these massive objects. 

ETGs exhibit smooth radial light distributions, which can be roughly approximated by a de Vaucouleurs profile \citep[e.g.][]{dV48,dV53,Kormendy77,Schweizer79}. However, ETGs are in general better described by S\'ersic profiles \citep{Sersic}, where a S\'ersic index n=4 corresponds to a de Vaucouleurs profile. Moreover, both n and effective radii \re \ of ETGs depend on galaxy mass or luminosity, with more massive galaxies having more concentrated light profiles (n$\gtrsim$4) and larger sizes \citep[e.g.][]{Graham03,Ferrarese06,Kormendy09,Kormendy12,Cappellari13b}. This change in the internal structure of ETGs with increasing mass is also observed in the shape of the isophotes, as low-mass galaxies show diskier contours, while the isophotes of more massive and luminous ETG tend to be boxier \citep{Bender89,Kormendy96,Kormendy12}. Moreover, more massive ETGs also tend to be more pressure-supported. On the contrary, rotating disks are common among low-mass ETGs \citep[e.g.][]{Emsellem07,Emsellem11,FB17,vdS17}. These morpho-kinematic characteristics \citep[see][for detailed reviews]{Graham13,Cappellari16} suggest that galaxy mergers, both minor and major, play a dominant role in shaping the observed properties of ETGs at redshift $z\sim0$ \citep[e.g.][]{Hibbard99,Naab06,Naab14,Burkert08,Johansson09,Duc11,navarro,Tapia14}, in agreement with the bottom-up nature of the $\Lambda$-CDM Universe.

The stellar populations in ETGs tell, however, a different story. Their stars are predominantly old, metal-rich and exhibit an enhancement of $\alpha$ elements with respect to Iron \citep[e.g.]{Peletier,Bender,vazdekis97,worthey92}. Both age, metallicity and [$\alpha$/Fe] in ETGs increase with galaxy mass \citep{Trager00b,Thomas05,Thomas11,Gallazzi05,Gallazzi06,cgvd}. Moreover, there is growing evidence of a non-universal, steep stellar initial mass function in the central regions of the most massive ETGs \citep[e.g.][]{MN15c,FLB16,Lyubenova16,vdk17,Oldham18}, transitioning towards Milky Way-like slopes in low-mass ETGs \citep[e.g.][]{treu,cappellari,FLB13,Spiniello12}. These tight scaling relations, also measured at high redshift \citep[e.g.][]{Bower92,Ellis97,vdk00,Blakeslee03,Bernardi05}, suggest that the bulk of the stellar populations within ETGs have been formed {\it in situ}, and not accreted over time. Hence, the so-called downsizing of ETGs is in tension with simplistic merger-driven scenarios for the formation of these objects.

Arguably, the most critical point in our understanding of galaxy formation is how the star formation in massive galaxies, and in particular in ETGs, is quenched. In the absence of additional mechanisms, the continuous accretion and recycling of gas leads to too many, too young massive galaxies in the local Universe, which is strongly rejected by observations \citep[e.g.][]{Burstein,Bressan96,Jorgensen99,vazdekis99,Trager00b,Terlevich02,McDermid15}. For the most massive objects, where environmental effects such as ram pressure or tidal stripping are negligible, three main channels have been proposed to explain the lack of star formation in massive galaxies.  First, gas heated through virial shocks in massive halos (i.e. {\it halo quenching}) has proved to be a successful mechanism to reconcile observations and theoretical expectations \citep[e.g.][]{Rees77,Birnboim03,Keres05,Dekel06,Bower06,Cattaneo08}. Alternatively, feedback from growing super-massive black holes could also heat up the gas in massive galaxies, quenching the star formation. This black hole feedback is widely adopted in state-of-the-art numerical simulations, and observational evidence suggest that it might actually drive star formation in massive objects \citep{dM05,Springel05b,Hopkins06,Sijacki07,Somerville08,Vogelsberger14,Schaye15,Weinberger17}. Finally, the formation of a compact bulge in the center of a massive galaxy could also stabilize the gaseous disk against star formation \citep{Martig09}. This so-called morphological quenching is observationally supported by the structural properties of massive galaxies at high redshift.

In order to reconcile all the observed properties of massive galaxies an emerging paradigm has been developed over the last years, rooted on both theoretical and observational results. In short, the so-called {\it two phase} formation scenario would consist on an initial gas-rich (dissipational) collapse onto dark matter halos, where the bulk of the stellar component of massive (ETGs) is formed \citep[e.g.][]{Oser,Oser12,navarro,Shankar13,Barro13,Pastorello14,MN15b,Foster16}. Regardless of the exact mechanism that drives the gas to the center of the massive halos (e.g. disk instabilities, accretion from the cosmic web or wet mergers), this initial phase leads to a dense, cuspy stellar core in a rapid formation process. This process takes place at high redshift ($z\gtrsim2$) and sets the overall properties of massive ETGs, which ultimately translate into the scaling relations observed in the local Universe \citep{Anna15,MN18c}. From redshift $z\sim2$ and below, the evolution of the most massive galaxies is driven by dry accretion of satellites and eventual (dry) major merger events. This final phase is thought to be responsible for the size growth of massive ETGs \citep{Daddi05,Trujillo07,vdk10b}, the change in their internal structure and kinematics \citep{Naab14,Cappellari16}, and the evolution of the stellar population gradients \citep{SB07,Spolaor10,MN18c}.

The success of this picture puts, however, unavoidable constraints to our observations. The light encoding information about high-$z$ baryonic processes is {\it contaminated} by the presence of accreted material along the line-of-sight. Moreover, the second phase not only hides the properties of the in situ-formed stars in massive galaxies, but it also affects their spatial and orbital distributions, irreversibly limiting our resolving power \citep[e.g.][]{Naab14}. Fortunately, the stochastic nature of the $\Lambda$-CDM Universe allows for the survival of galaxies that, after the initial collapse, have remained unaffected by significant merger episodes \citep{Taylor10,Quilis13,Stringer15}. These objects, commonly referred as naked red nuggets or massive relic galaxies, are local analogs to the high-$z$ population of galaxies, and they offer a unique opportunity to study in great detail the galaxy formation processes and the physical conditions of the early Universe. A ten-year effort to find this precious population of galaxies \citep[e.g.][]{Trujillo09,Taylor10,Valentinuzzi10,Poggianti13,Saulder15,Tortora16} has finally lead to the discovery of a handful of nearby galaxies fulfilling all the required conditions to be considered naked red nuggets \citep{Remco12,Trujillo14,MN15b,akin17,Anna17,Beasley18}. 

In this work, we explore the sample of naked red nuggets presented in \citet{akin17}. The basic stellar population properties of these objects have been extensively described in \citet{akin17} and in \citet{MN18c}. Here, we develop a complementary photometric and stellar population analysis based on the measured star formation histories (SFH) of our sample. The basic idea is simple: we decompose the optical spectra into a series of single stellar population models with varying age. In this way, we can measure the amount of light and mass that is contributed by the different stellar populations present within a galaxy. We then combine the measured SFHs with the two-dimensional spatial information provided by integral field spectroscopy (IFU). In practice, this combination allows us to trace back the structural properties of individual galaxies as a function of age, redshift, or any other stellar population parameter. Although observationally expensive, our approach can be applied to any object observed through an IFU and presents three main advantages. First, it is by construction insensitive to progenitor biases. Second, it allows a direct comparison of galaxies at the same evolutionary stage, regardless of the intrinsic redshift of the objects. Third, it is a powerful tool to probe the early Universe with a precision level only achievable at lower~$z$. 

The layout of this work is as follows. In \S~2 we introduce our sample and data. In \S~3 we describe the stellar population analysis, and our results are presented in \S~4. A discussion about the quenching process is presented in \S~5, and summarized in \S~6. 

\section{Sample and data}
Our analysis is based on the integral-field unit spectroscopic data presented in \citet{akin17}. In short, a sample of 12 naked red nuggets was observed with the Potsdam Multi Aperture Spectrograph (PMAS; Roth et al. 2005) and the PPAK fiber module (Verheijen et al. 2004; Kelz et al. 2006), with the V500 grating. This setup effectively provides a $\sim$40''x40'' field-of-view, with a spatial resolution of 1''. The spectral coverage ranges from $\sim3700$\AA \ to $\sim$7300\AA, with a fairly constant spectral resolution of 6\AA \ (FWHM) across the whole wavelength range. Additionally, $HST$ $H$-band photometry has been acquired for the whole sample \citep[see][\S2.1]{akin17}.

The 12 naked red nuggets expand a range in stellar mass from $\log M_\star / \msun =9.89$\ to  $\log M_\star / \msun =11.34$, with typical sizes of \re$\sim$1.5 kpc. They are all nearby ($D\lesssim$~100~Mpc) and commonly located in dense environment \citep[see][]{Peralta16}. More details on the sample properties can be found in \citet{akin17}.

Since PMAS-PPAK data are flux calibrated, the IFU data cubes can be collapsed along the spectral dimension into a two-dimensional image, with an effective photometric filter given by a boxcar kernel across the (observed) wavelength range.

\section{Stellar population analysis}

In order to do the structural analysis of our sample, we wanted to preserve as much spatial information as possible, while having at the same time enough signal-to-noise to reliably measure SFHs. Thus, we performed a Voronoi-binning \citep{voronoi} tessellation of the IFU field-of-view, imposing a minimum signal-to-noise ratio of 60 per \AA.

We measured SFHs using the STEllar Content and Kinematics via Maximum A Posteriori likelihood (STECKMAP) code \citep{Ocvirk06,Ocvirk06b}, which finds the best linear combination of single-age stellar population models. For each of these coeval models, the metallicity is allowed to vary. Since finding the best-fitting combination of models is an ill-defined problem \citep[e.g.][]{Cappellari17}, STECKMAP adopts a regularization scheme for both age and metallicity distributions. The amount of regularization is determined by the (dimensionless) parameters $\mu_x$ and $\mu_Z$, where $\mu_x$ applies to the age distribution and $\mu_Z$ regularizes across the age-metallicity relation. We imposed  $\mu_Z = \mu_x = 10$, and we checked that this choice does not affect our analysis. Note that we first measured the kinematics using pPXF \citep{ppxf}, and then we kept it fixed for the stellar population analysis. This strategy has shown to minimize the velocity dispersion-metallicity degeneracy \citep{Koleva08,Pat11}. Errors in the analysis were calculated using 25 Monte Carlo simulations based on the 1-$\sigma$ uncertainties of the Voronoi-binned spectra. We limited the fitting to wavelengths in between 3750\AA \ and 5400\AA.

We fed STECKMAP with the MILES stellar population synthesis models \citep{miles,milesII}, based on the MILES stellar library \citep{milesPAT}. We use the so-called {\it base} MILES models, which follow the solar neighborhood abundance pattern, with the Padova set of isochrones \citep{Girardi00}. These base models cover a range of metallicities from [M/H]= $-2.32$dex to [M/H] = $+0.26$ dex, and a range of age from $\sim$0.06 to 17 Gyr. Ages older than the age of the Universe are allowed due to well known model systematics affecting the temperature scale of the isochrones \citep{vazdekis01,Schiavon02}. In particular, the effect of non-solar abundances \citep{vazdekis01} and non-universal IMF \citep[e.g.][]{FLB13} would artificially shift inferred ages towards older values. Since galaxies on our sample are known to be [Mg/Fe] enhanced and with a bottom-heavy IMF \citep{MN15b,MN18c}, the upper limit of 17 Gyr was allowed for safety. Absolute age values are therefore not reliable, and age differences should be considered in relative terms.

It is worth noting that STECKAMP has been extensively tested. \citet{RL15,RL18} found that the STECKAMP-based SFHs are in great agreement with those measured using color-magnitude diagrams. In addition, STECKMAP has been used to measure the stellar population properties of a wide variety of objects, from spiral galaxies \citep{Pat11} to globular clusters \citep{Ocvirk10}, showing consistency with similar full spectral fitting codes \citep{Koleva08}. Moreover, \citet{Ocvirk06} demonstrated that the STECKAMP ability to separate different age components mostly depends on the signal-to-noise ratio of the data, while the sensitivity on other parameters such as spectral resolution is much weaker. The spectral range given by our instrumental set-up is similar to that used by \citet{Ocvirk06} to study the robustness of STECKAMP. With a signal-to-noise ratio of 60 per \AA, the accuracy of our results is then backed up by both theoretical \citep{Ocvirk06,Ocvirk06b} and observational \citep{RL15,RL18} results.

We notice, however, that as any other analysis of unresolved stellar populations, STECKAMP ages and SFHs are partially degenerated with the metallicity determination \citep[e.g.][]{Worthey94b}. This intrinsic limitation can be particularly relevant for naked red nuggets since they tend to exhibit extreme chemical composition patterns \citep{MN15b,MN18c} and the stellar library feeding the MILES models \citep{milesPAT} does not contain stars with such an unusual chemistry. This extreme chemistry also makes the comparison with resolved stellar population studies less direct as the latter have generally studied less metal-rich systems \citep[see e.g.][]{RL15,RL18,Ocvirk10}. Moreover, as described above, we allowed in our analysis for stellar populations older than the age of the Universe to account for model systematics \citep[e.g.][]{Schiavon02}. With all these caveats in mind, the results presented in this paper should be always interpreted in a relative manner and not as absolute values.

\begin{figure*}
\begin{center}
\includegraphics[height=8.cm]{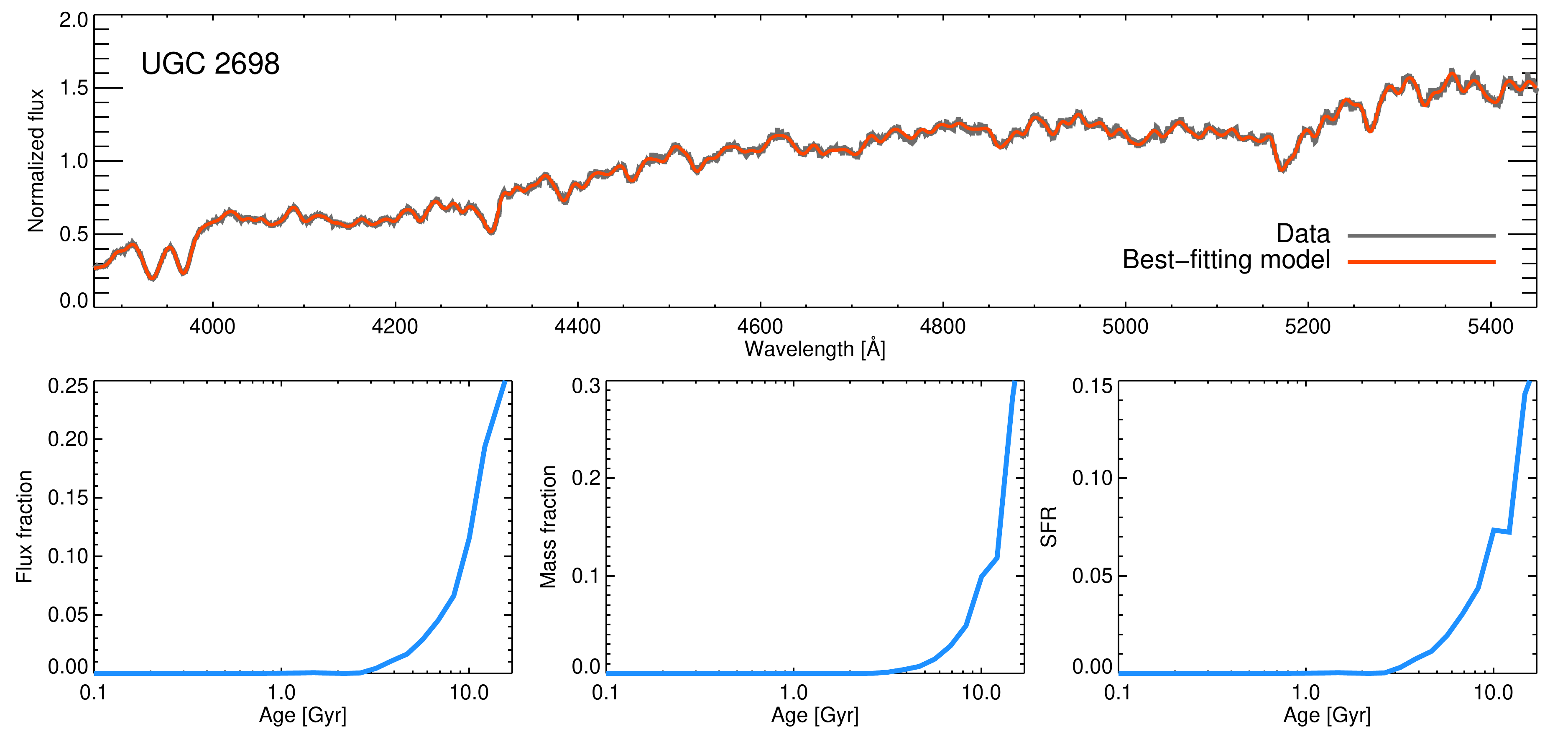}
\end{center}
\caption{The upper panel shows, in black, the data from the central spaxel of UGC\,2698, and in red the best-fitting model from STECKMAP. This best-fitting model consists of a linear combination of single stellar population model from the MILES set of templates (bottom left). Knowing the mass-to-light ratio of each model, this best-fitting linear combination can be translated into mass fraction as a function of the age of the model (bottom center). Finally, combining the mass fraction and the time-step of the stellar population models, the observed spectrum of a galaxy can be interpreted as a function of a time evolving star formation rate. (bottom right). We recover these three quantities (flux ratio, mass ratio and SFR) over the whole field-of-view of the spectrograph, allowing us to dissect in time and space the evolution of massive red nuggets.}
\label{fig:products}
\end{figure*}

In summary, for each Voronoi-binned spectrum STECKAMP measures, in a non parametric way, the SFH and the age--metallicity relation of the underlying stellar populations \citep[see][for details]{Ocvirk06}. In this way we recover three quantities: the flux, the mass, and the star formation rate (SFR) as a function of the age of the populations. This is exemplified in Fig.~\ref{fig:products}, where we show the central spectrum of UGC\,2698, the best-fitting model, and the byproducts of the STECKMAP analysis.

\section{Results}

Having the STECKAMP-based SFH maps, we studied two main properties of our sample of galaxies. First, we investigated the relation between specific star formation rate and central density. Second, we explored how the surface brightness profiles of naked red nuggets evolves in time.

\subsection{Quenching}

\begin{figure}
\begin{center}
\includegraphics[width=8.5cm]{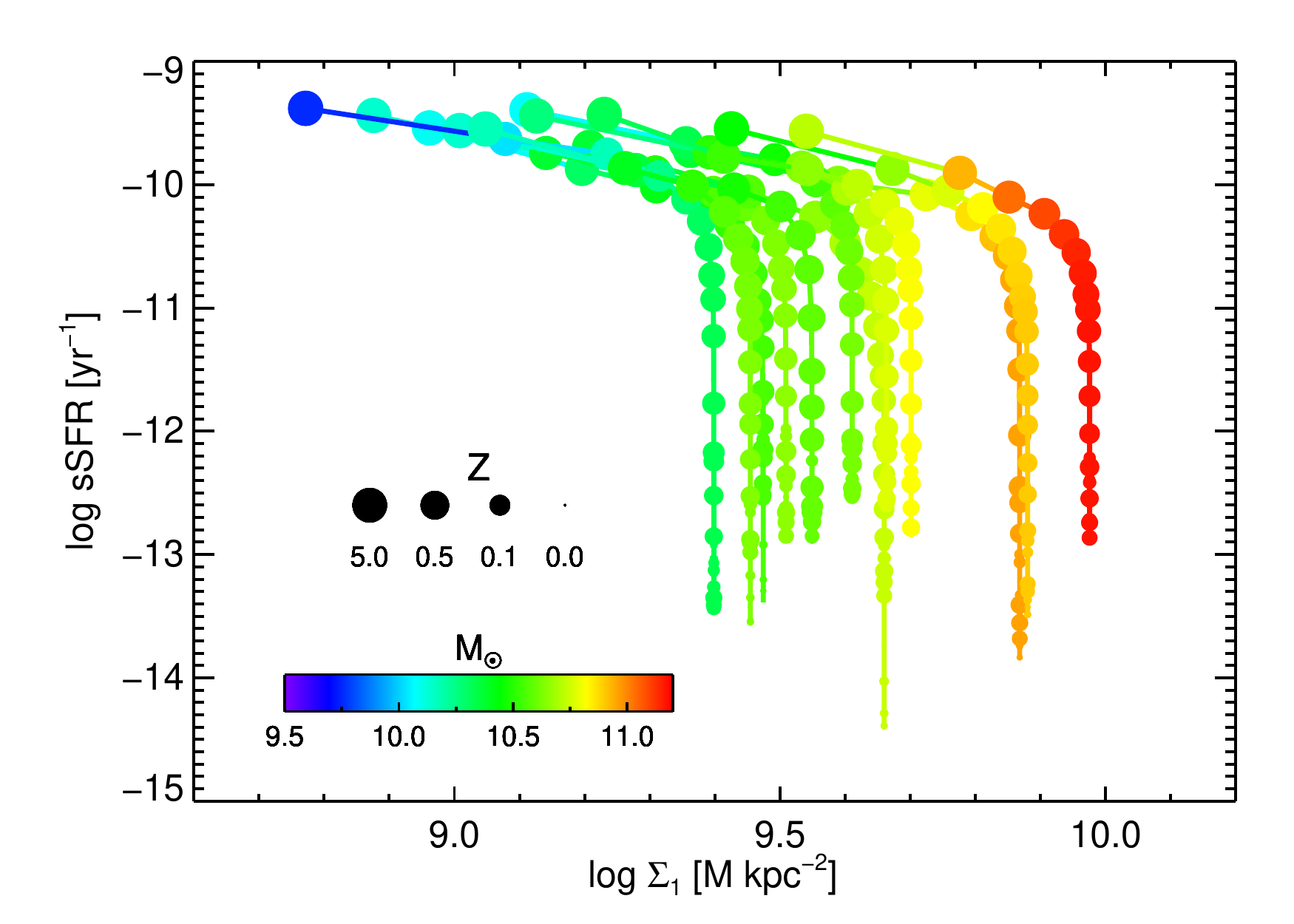}
\end{center}
\caption{Specific SFR vs central density. Solid lines show the evolution of each naked red nugget in our sample, color-coded by the $\log$ of the stellar mass of the galaxy at that time. Symbols indicate the age/redshift of each stellar population. Galaxies start with high SFRs and low central densities (top left corner). The quenching of the star formation corresponds to the well-defined knee in these evolutionary tracks. The quenching process is fast but it is not uniquely defined by a single central density $\Sigma_1$ value. Instead, a clear trend is visible, with more massive galaxies quenching at higher $\Sigma_1$.}
\label{fig:quench}
\end{figure}

The quenching history of our sample of naked red nuggets can be explored using the measured SFHs. In particular, it has been proposed that massive galaxies follow well-defined tracks across the specific star formation rate (sSFR) -- central stellar mass density plane \citep[e.g.][]{Cheung12,Barro13,Whitaker17}. The use of SFHs enables, although not entirely free of systematic effects (e.g. template mismatch, PSF smearing, fitting degeneracies), a more direct comparison to theoretical predictions, since the time evolution of individual objects can be explored. Thanks to the spatially-resolved SFHs, both SFR and total stellar mass of a galaxy are known for any given look-back time or redshift.  sSFR was measured as the averaged SFR of a galaxy within 3\re, divided by the stellar mass within the same aperture. Similarly, we traced back the time evolution of the stellar mass density within 1 kpc ($\Sigma_1$) by calculating the SFH as measured from the average absorption spectrum of the central 1 kpc of each galaxy. In Fig.~\ref{fig:quench} we show the measured sSFR-$\Sigma_1$ tracks for our sample of massive compact galaxies.

The similarity of these tracks to the theoretically predicted is remarkable \citep[see Figure 6 in][]{Kocevski17}. While the vertical axis probes the overall sSFR of each object as a function of time, the horizontal axis shows the growth of the central component. Color indicates the stellar mass within 3\re \ at every time step, while time evolution is represented by symbol size. It is clear from Fig.~\ref{fig:quench} that $\Sigma_1$ steadily increases with time and reaches a maximum that coincides with the overall quenching of the galaxy. The characteristic $\Sigma_1$ value at quenching appears to be a function of stellar mass as more massive galaxies get quenched at higher $\Sigma_1$.

As noted above, the absolute values of our SFHs might be partially degenerated with our metallicity determination. However, the trends shown in Fig.~\ref{fig:quench}, and particularly the sudden drop in the SFRs, ultimately reflect the fact that the SFHs of massive relic galaxies declined with time, which is independent of the assumed metallicity. In addition, since more massive galaxies are expected to be more metallic and therefore redder, the age-metallicity degeneracy would artificially shift our most massive objects towards younger (bluer) ages, which is the opposite to what is observed in Fig.~\ref{fig:quench}.

\subsection{Stellar population-based photometry}

\begin{figure}
\begin{center}
\includegraphics[width=8.cm]{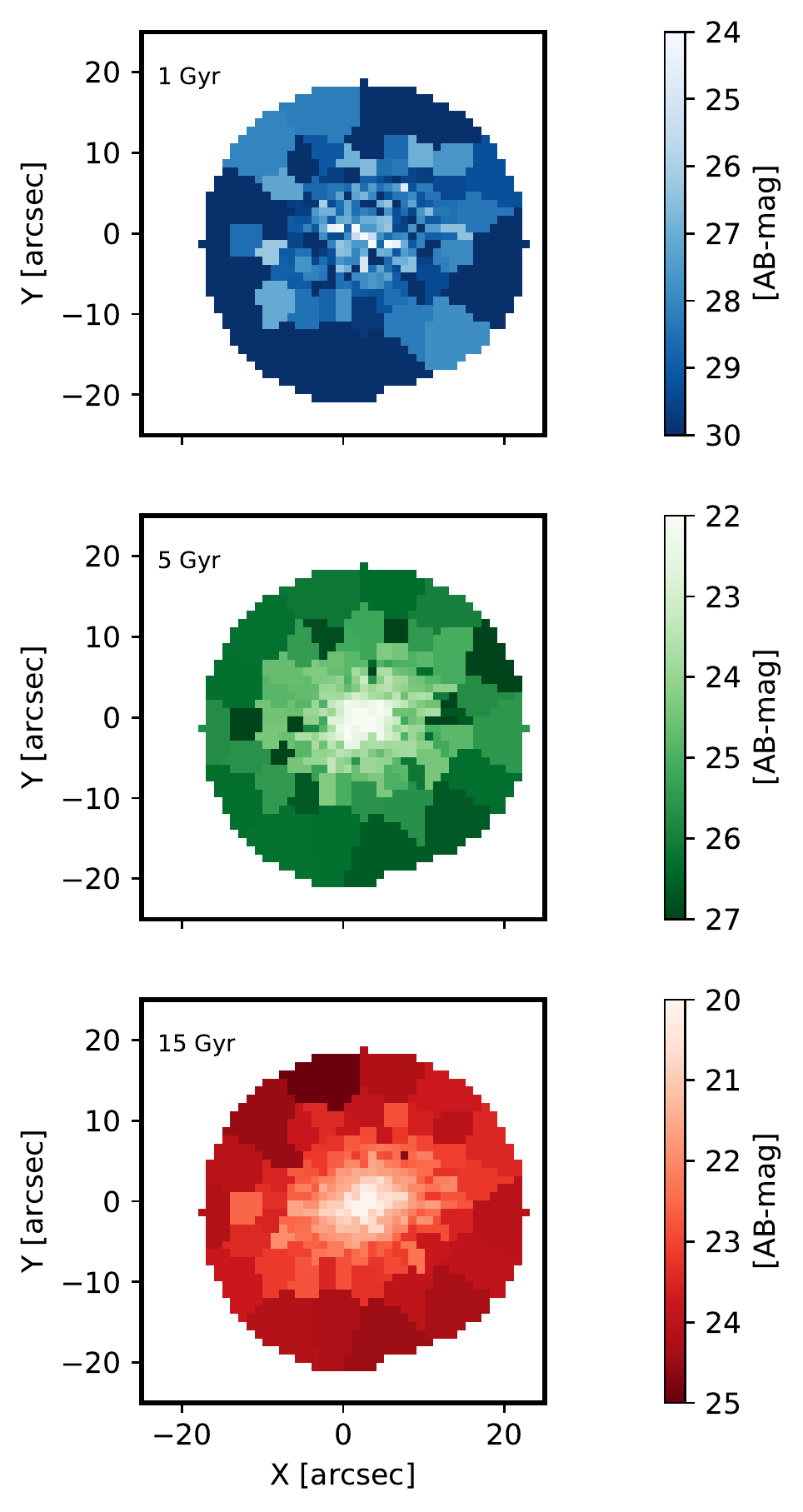}
\end{center}
\caption{Distribution of the stellar light of UGC\,2698 formed at different times. From top to bottom, spatial distribution of the stars that were formed 1 Gyr, 5 Gyr, and 15 Gyr ago, respectively. Color bars indicate the effective surface brightness, which becomes weaker for younger stellar populations. Notice that the scale in the color bars changes from panel to panel in order to compensate the fact that younger populations contribute less to the integrated flux of the galaxy. It is clear form this figure that the internal structure of massive relic galaxies changed with time as stars formed under different conditions.}
\label{fig:ages}
\end{figure}

One of the main advantages of coupling SFHs to IFU data is that we can recover the structure evolution of individual objects as a function of time, assuming that stellar redistribution processes such as stellar radial migration or heating are subdominant. As an example, we show in Fig.~\ref{fig:ages} the spatial distribution of the stellar populations in UGC\,2698 formed at three different times. Another important point to be highlighted is that every Voronoi bin is decomposed by STECKMAP independently of the others. In practice, this means that the observed radial trends are free from systematics in the stellar population modeling.

For every object, the surface brightness profile was calculated by collapsing the IFU cubes along the spectral dimension. To be consistent with the stellar population modeling, we only collapsed flux within the fitted wavelength range, i.e., between 3750\AA \ and 5400\AA. We adopted distances to the center of the galaxy, position angles and ellipticities from \citet{akin17}. Unfortunately, the SN of the pseudo-light profiles was not enough to perform a robust S\'ersic fit for individual galaxies (see top panel in Fig.~\ref{fig:ages}). Therefore, we combined all our sample into a single pseudo-galaxy, scaling the profiles to an common effective radius of 2 kpc. Since the effective flux coming from young populations was very low (see blue symbols in Fig.~\ref{fig:ages}), we further binned the data into eight different age bins.

\begin{figure}
\begin{center}
\includegraphics[width=8.cm]{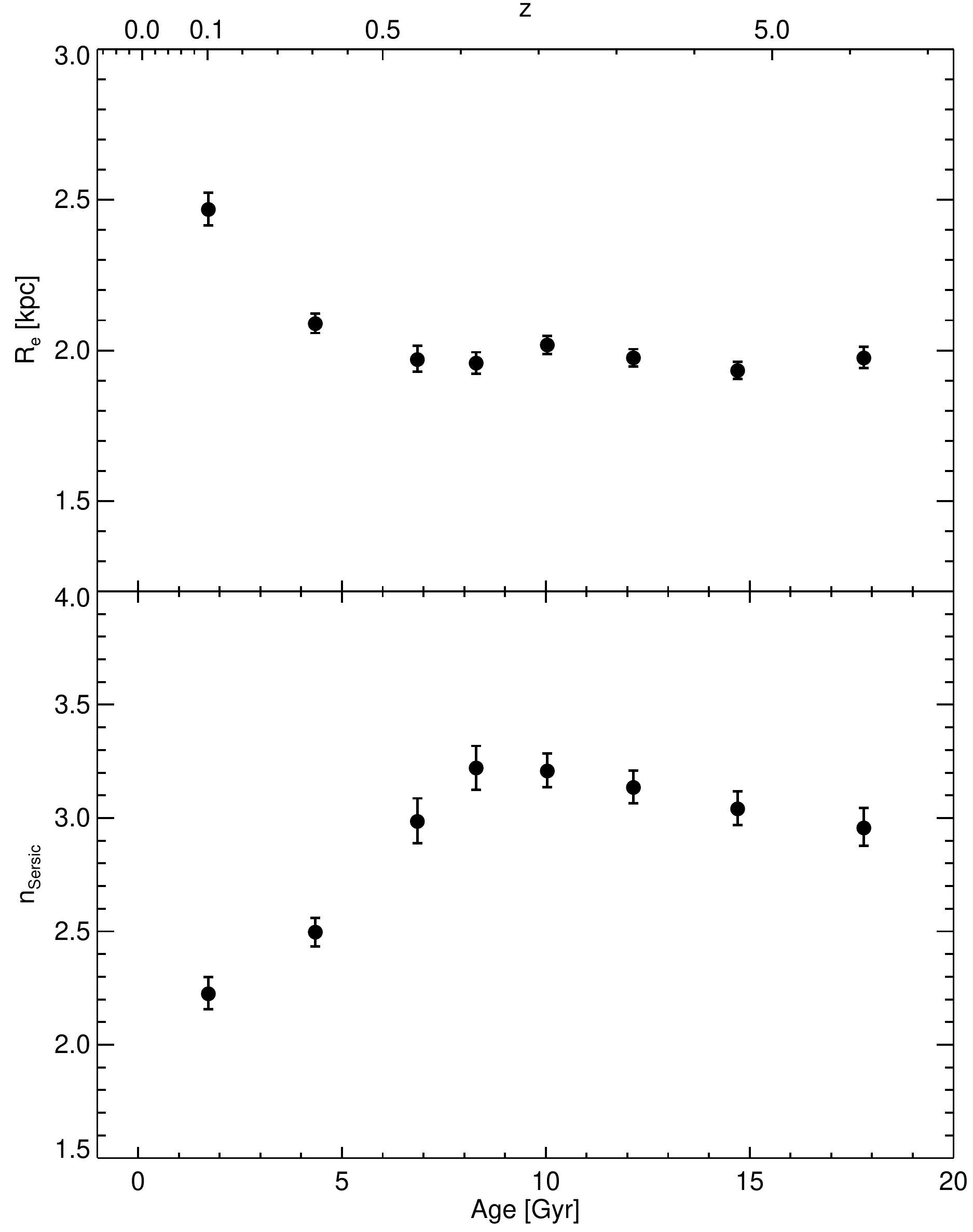}
\end{center}
\caption{Evolution of the morphological parameters. A S\'ersic fit was not possible in individual galaxies and here we show the average behavior of all our sample. ({\it Top}) Effective radius as a function of age of the stellar populations. For ages older than $\sim$8 Gyr, the effective radius is rather constant, but it becomes larger for younger stars. ({\it Bottom}) Same as above but for the S\'sersic index $n$. In this case, $n$ reaches a maximum around  $\sim$8 Gyr ago, decreasing for older and younger stellar populations. Although these morphological parameters (top and bottom panels) were independently measured for stellar populations with different ages, the flux of our galaxies, and therefore their stellar masses, are entirely dominated by old stars (see Figs.~\ref{fig:products},~\ref{fig:ages}).}
\label{fig:param}
\end{figure}

We fit these eight surface brightness profiles with a S\'ersic model using the {\it emcee} Bayesian Markov chain Monte Carlo sampler \citep{emcee}. We used flat priors for the three free parameters, namely, central surface brightness ($\mu_0$), effective radius (\re) and S\'ersic index ($n$). The assumed photometric uncertainties include both photometric and stellar population errors. Since we scaled the flux of individual galaxies to account for differences in (apparent) brightness among them, the only meaningful parameters are effectively \re \ and $n$, which are shown as a function of age in Fig.~\ref{fig:param}.

The effective radius of naked red nuggets remains roughly constant in time for populations older than $\sim5$ Gyr, and then it tends to increase towards younger populations. The time evolution of the S\'ersic index on the other hand is more complex. Very old stars show relatively low S\'ersic index values, but $n$ increases with time, peaking at around 8 Gyr. A null age--$n$ relation for stars older than 8 Gyr is rejected at a $3.3\sigma$ level. For younger stellar populations, the light profile becomes less concentrated (i.e., low $n$) again. Note that error bars correspond to 1$\sigma$ confidence intervals of the posterior distributions. 

\section{Discussion}

\subsection{Morphological trends: two modes of star formation}

In nearby naked red nuggets stellar populations with different ages show different radial distributions (Fig.~\ref{fig:param}).  In particular, we have shown how the effective radius remains almost constant up until $\sim8$ Gyr ago, but then it increases for younger ages. Note that individual points in Fig.~\ref{fig:param} are not integrated values, i.e., they do not show how the galaxy would look at that time, but how the stars born at that time are distributed. The observed trend in \re, increasing with decreasing age of the stars, is effectively equivalent to the larger sizes of galaxies observed through bluer photometric bands \citep[e.g.][]{Evans94,Cunow01,Mollenhof06,Kelvin12} as younger stars will dominate the flux at shorter wavelengths.

The peak in the S\'ersic index evolution reveals a more complex scenario. The bottom panel in Fig.~\ref{fig:param} suggests that a compaction phase takes place during the early live of red nuggets. This initial compaction is a well-known theoretical expectation, where gas is efficiently channelled towards the center of dark matter halos, leading to massive, gas-rich, compact, and star-forming disks \citep[e.g.][]{Gammie01,Dekel09,Dekel13,Dekel14,Dekel19,Krumholz10,Cacciato12,Zolotov15}. Observations of massive, compact and star-forming galaxies at high-$z$ support this scenario \citep[e.g.][]{Barro13,Barro14,Barro17,HC15,vdk15}, but linking this population of galaxies to their more extended progenitors is not trivial. Our approach, tracing back in time the structural evolution of individual objects, shows that for our sample of galaxies an initial compaction took place, reaching a maximum $\sim$8~Gyr ago ($z\sim1$).

\begin{figure}
\begin{center}
\includegraphics[width=8.cm]{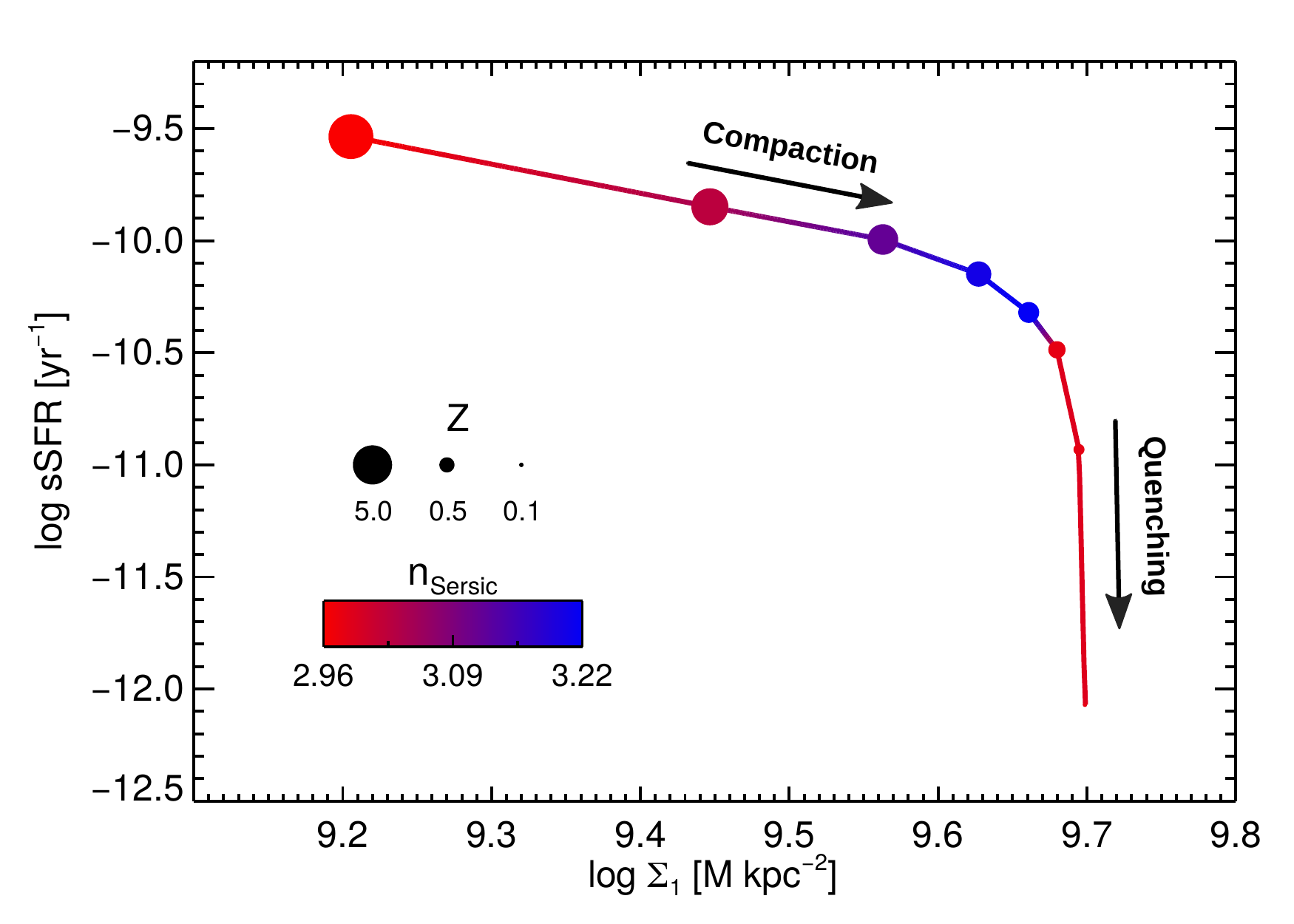}
\end{center}
\caption{Morphology -- quenching relation. The average SFR vs $\Sigma_1$ for our sample of naked red nuggets is shown, color-coded by the S\'sersi index of the stellar populations at that time. The clear connection between SFR and morphology suggests two modes of star formation: while star forming, red nuggets become more concentrated in time, leading to high S\'ersic index and $\Sigma_1$ values. The quenching process takes place at the highest $n$ value. The residual star formation during the quiescent phase happens in a more disk-like structure.}
\label{fig:compa}
\end{figure}

In order to explore the connection between SFR and compaction, we show in Fig.~\ref{fig:compa} the average evolution of sSFR--$\log \Sigma_1$ in our sample, color coded by the S\'ersic index at that time (from Fig.~\ref{fig:param}). A clear connection can be seen between star formation activity and galaxy structure. The initial compaction takes place during the star-forming phase, as expected in a gas rich environment. Later, the actual quenching takes place when the compaction is maximal. This could be naturally explained if the gas brought towards the center or the galaxy efficiently feeds the central super-massive black hole. The black hole would become therefore energetic enough to shut down star formation, likely leading to an inner bulge-like component \citep[e.g.][]{Silk98,King,dM05,Sijacki07,Silk13}. Note also that the build up of the central component is likely associated with a strong stellar feedback even, which would also favor the quenching process. The effect of stellar feedback might be emphasized if the IMF is flatter (i.e. dominated by massive stars) during this process \citep{Vazdekis96,weidner:13,Zhang18}. Once the stellar component is compact enough, morphological quenching could also prevent further star formation \citep{Martig09}. This relation between star formation and morphology is in agreement with previous studies showing statistical trends among large samples of galaxies at different evolutionary stages \citep[e.g.][]{Bell12,Bluck14,Lang14,Omand14,Mosleh17,Wang18}. The quenching is followed by a final phase characterized by a low level star formation (see also Fig.~\ref{fig:ages}), where stars form in a more relaxed, disk-like structure. 

It is worth highlighting that the quantities presented in Fig.~\ref{fig:compa} are highly decoupled. While the structural analysis is based on the light decomposition across spaxels whose SFHs have been independently calculated, the sSFR depends on the evolution of the overall star formation, and $\Sigma_1$ is driven by the very central regions of our galaxies. Our stellar population-based photometry is, however, heavily limited by the large point-spread-function (PSF) of the PMAS-PPAK spectrograph which in practice means that the SFHs of adjacent spaxels are correlated. Therefore, the absolute S\'ersic index values and R$_e$ shown in Figs.~\ref{fig:param},~\ref{fig:compa} should be only considered as upper limits. Since we do not have stars bright enough around our massive red nuggets, a proper characterization of the PSF was not possible. Without a proper PSF model deconvolving the (noisy) pseudo-images (Fig~\ref{fig:ages}) was not feasible. The trend shown in Fig.~\ref{fig:compa} is however har to explain by invoking an instrumental bias as there is no reason for the PSF to depend on the age or on the SFR or the underlying stellar populations. Therefore, and despite of these systematics, our analysis suggests a clear connection between star formation activity  and morphology, suggesting two modes of star formation: an initial highly star forming phase characterized by a structural compaction, followed by a quenched stage, where stars form in a more radially extended way. 

\subsection{Time evolution of black hole scaling relations}

It has been long debated if the observed scaling relations between black hole mass and galaxy-wide properties evolve over time \citep[e.g.][]{Treu04,Walter04,Robertson06,Merloni10,Volonteri10,Lamastra10,Taylor16,Saito16,Volonteri16,Ding17,Huang18}. A variety of mechanisms have been proposed and calibrated to explain the emergence of these scaling relations at $z\sim0$, with different physical processes leading to different redshift evolutions \citep[e.g.][]{Yu04,Robertson06,Jahnke11}. Hence, observationally constraining this time dependence is key to understand, not only the growth of super-massive black holes, but also how (if) they regulate baryonic cooling in their host galaxies. 

Direct black hole mass measurements are however observationally expensive, and technically impossible beyond $z\simeq0.5$ even for the next generation of 30-meter telescopes. Alternatively, locally-calibrated proxies based on the properties of the accretion disks around growing black holes have been used to explore scaling relations at higher redshifts \citep[e.g.][]{Treu04,Walter04,Jahnke09,Merloni10,Decarli10,Bluck11,Shen15,Saito16}. 

\begin{figure}
\begin{center}
\includegraphics[width=8.cm]{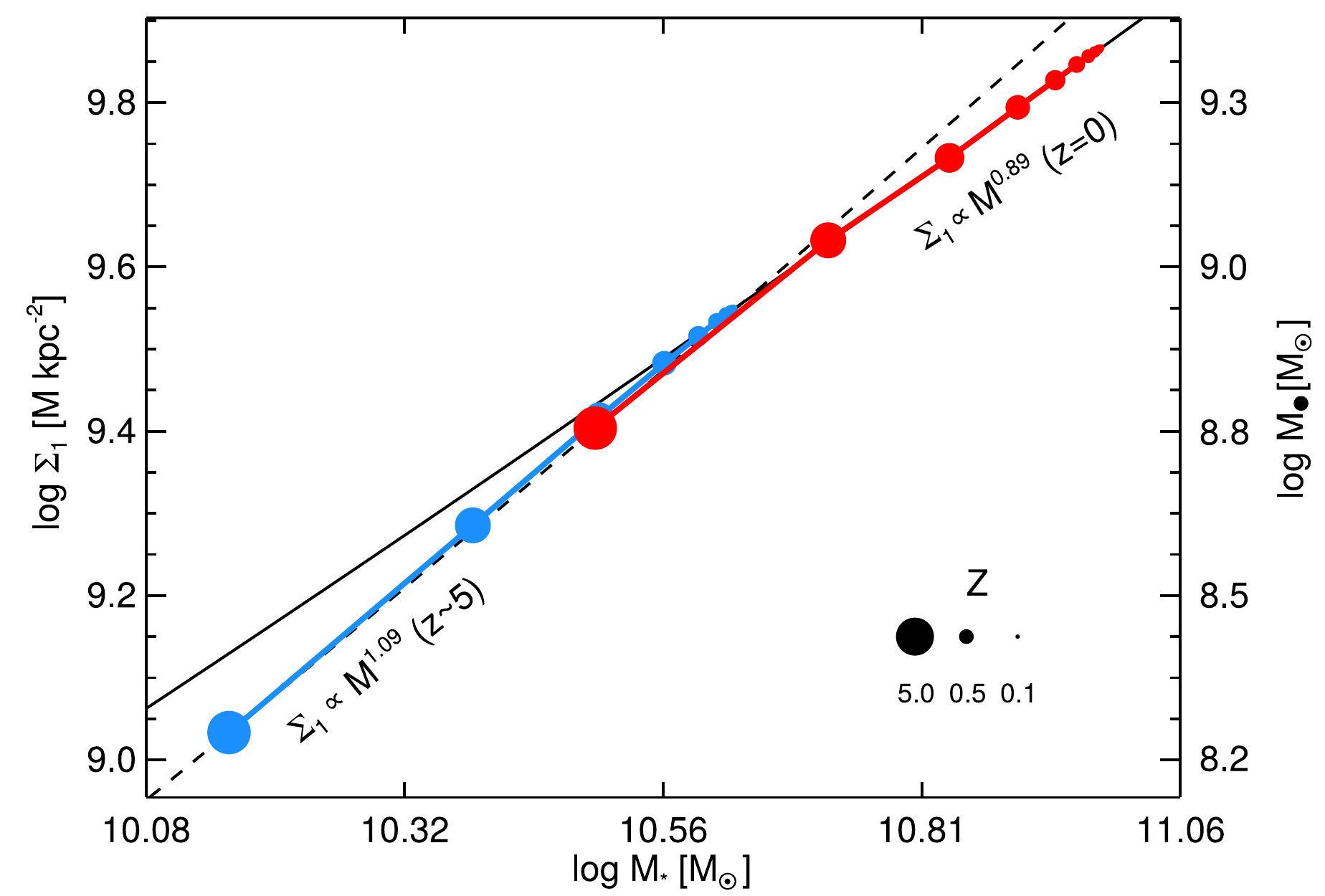}
\end{center}
\caption{Central density and galaxy mass growth. Evolution of the central density $\Sigma_1$ as a function of stellar mass. Blue and red curves show the evolutionary track of high-mass and low-mass naked red nuggets in our sample. An initial phase where the $\Sigma_1$ grows faster than the stellar mass is clear, in particular for lower-mass objects. This figure suggests an initial phase of more rapid black hole growth, followed by a dominant (in time) phase were black holes and galaxies grow at a similar pace. For the massive sub-sample, the initial phase was likely to rapid to be resolved from $z\sim0$. These tracks suggest that black hole scaling relations have been fairly constant in time (black solid line), except from an brief initial stage at high redshift ($z\sim5$, dashed line).}
\label{fig:evol}
\end{figure}

Our stellar population analysis allows us to propose here an alternative observational approach to understand the evolution of black hole scaling relations. In Fig.~\ref{fig:evol} we show the mean evolutionary tracks for our sample of naked red nuggets across the $\log \Sigma_1$ and $\log M_\star$ plane. Two separate lines are shown, in red for massive galaxies (more massive than the mean stellar mass of the sample $\log M = 10.78$ \msun) and in blue for lower mass objects. Each point correspond to a different look-back time (or redshift). Interestingly, the evolution of galaxies in this plane is mass-dependent. More massive objects in our sample evolved along a line of roughly constant $\log \Sigma_1$-to-$\log M_\star$ ratio. Lower mass objects on the other hand experienced an initial phase where $\log \Sigma_1$ grew more rapidly  than the total stellar mass (e.g. galaxies moved rather vertically in the $\log \Sigma_1$--$\log M_\star$ plane), and then they settled and evolved along the $z\sim0$ relation. 

In practice, Fig.~\ref{fig:evol} can be used to probe the evolution in time of the black hole mass -- galaxy mass scaling relation. At redshift $z\simeq0$, black hole masses in our sample can me estimated via the velocity dispersion of the host galaxy \citep{vdb}. Assuming that black hole growth tracks the growth of the stellar mass within $\sim$ 1 kpc, Fig.~\ref{fig:evol} suggests that the relation between stellar and black hole mass steepens with redshift as it takes more time for low-mass galaxies to reach the $z\sim0$ characteristic ratio. Note that this change in the slope is only noticeable for look-back times larger than 10 Gyr, or equivalently, for redshifts $z\gtrsim2$, when it changes more abruptly \citep{Rosas16}. The validity of using the central stellar mass as a proxy for black hole growth will be tested in an upcoming work.

The evolutionary tracks in  Fig.~\ref{fig:evol} can be interpreted as two different modes of black hole and galaxy growth. Initially, black holes would experience a rapid (non-linear) growth, somewhat decoupled from the growth of the galaxy itself, which would move galaxies vertically in the $\log \Sigma_1$--$M_\star$ plane as explained above. According to Fig.~\ref{fig:evol}, this initial black hole growth would occur earlier in more massive galaxies, while lower mass galaxies would lag behind, leading to the observed steepening at high $z$. The subsequent growth of both black hole and galaxy mass would be self-regulated \citep[e.g.][]{Silk98,dM05,Robertson06,Booth09,Booth11,Mullaney12,Sijacki15}, along the $z\sim0$ track of constant $\Sigma_1$-$M_\star$ (solid line in Fig.~\ref{fig:evol}).

As any other indirect black hole mass estimation, our approach relies on the assumption that black hole mass growth tracks the growth of the central stellar component, which is what we measure from the integrated spectra. The physical motivation behind this assumption is that stars in our sample were formed in-situ. Thus, the same gas supply that lead to the formation of these stars must have fed the central super-massive black holes, increasing their masses proportionally to the growth of the stellar component. The main limitation of this approach is that it relies on a constant stellar-to-black hole mass conversion factor, which in principle might change as galaxies evolve. Aperture and merger effects \citep[e.g.][]{Faber97,Anna15} further complicate a clean interpretation. Therefore, the tracks shown in Fig.~\ref{fig:evol} should be interpreted as relative trends, keeping in mind that what we are actually measuring is the evolution of the central mass. Conversely, scales of $\sim 1$kpc are  already resolved in cosmological numerical simulations \citep{Vogelsberger14,Schaye15}. Hence, understanding the growth of black holes via integrated spectroscopy over the central regions of galaxies allows a cleaner comparison with observations, rather insensitive to sub-grid physical processes. This synergy between observations and theoretical predictions offers a unique opportunity to study galaxy and black hole mass growth well beyond the local Universe.

Finally, it is worth mentioning that our findings are not in contradiction with previous claims of a higher normalization of the M$_\bullet$--$M_\star$ relation with increasing redshift \citep[e.g.][]{Treu04,Walter04,Merloni10}. The key property of naked red nuggets is the fact that they have evolved {\it in isolation}, without significant merger events. The accretion of galaxies would mostly increase their mass over time, effectively moving naked red nuggets towards the right in Fig.~\ref{fig:evol} \citep[e.g.][]{Trujillo07,vdk10b,Anna17}. This process would lower the M$_\bullet$--$M_\star$ normalization over time.

\subsection{Black hole plus halo quenching?} \label{sec:bhalo}

Our analysis, and in particular the time variation of the star formation history, offers a complementary view to the quenching process of galaxies based on the evolution of individual objects. From Fig.~\ref{fig:quench} it is clear that there exists a dependence with stellar mass, where objects that at $z\sim0$ are more massive getting quenched at higher central densities. The trend between $\Sigma_1$ at quenching and galaxy mass could be understood as a cosmological effect. More massive galaxies could have formed earlier, when the Universe was denser, leading to higher $\Sigma_1$ \citep[e.g.][]{wellons}. We tested this scenario by comparing quenching time (defined as the knee in Fig.~\ref{fig:quench}) and galaxy mass in our sample. We find no significant correlation between these two quantities, with more massive galaxies quenching at slightly later times. Thus a cosmologically-induced compaction is not responsible for the scatter observed in Fig.~\ref{fig:quench}.

\citet{Bower16} presented a suggestive result from the EAGLE simulation, where star formation rate at fixed halo mass is linked to the mass of the central black hole. In particular, quiescent galaxies host more massive black holes than star forming galaxies with the same halo mass. The relatively isolated evolution of naked red nuggets can be used to investigate the black hole -- halo mass -- baryonic cooling connection. Star formation in red nuggets can be assumed to evolve at constant dark matter halo mass, since the accretion of material responsible for the halo growth has been suppressed. Red nuggets would have evolved as isolated dark matter halos filled up with gas. Under this assumption, we used the stellar mass -- halo mass relation (SMHM) of \citet{Behroozi13} to infer the mass of the dark matter halo in our sample. 

\begin{figure}
\begin{center}
\includegraphics[width=8.cm]{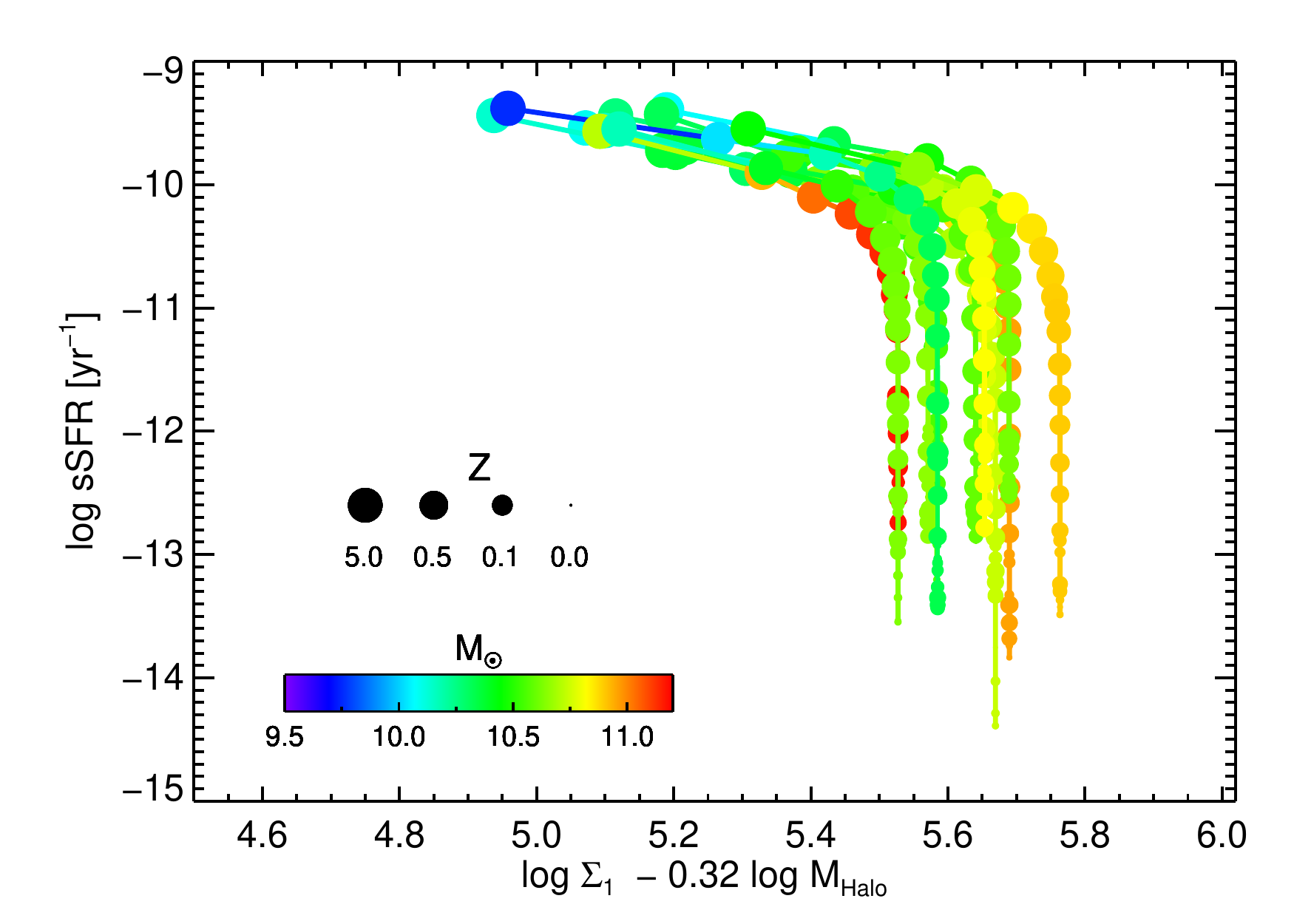}
\end{center}
\caption{Specific star formation rate as a function of central density and halo mass. Same as in Fig.~\ref{fig:quench} but as a function of a linear combination of $\log \Sigma_1$ and $\log M_{halo}$.  The coefficient was calculated to minimize the scatter in $\log \Sigma_1 + \beta \log M_{halo}$ at $z\sim0$. Halo masses are based on \citet{Behroozi13} ($z=2$) assuming an isolated evolution for naked red nuggets. Contrary to Fig.~\ref{fig:quench}, all tracks in this figure show a well defined quenching scale at $\log \Sigma_1 + 0.32 \log M_{halo} \simeq 5.6$, suggesting a combination of black hole and halo quenching. Galaxies with lighter black holes or halos would be able to sustain star formation, while galaxies with black holes and halos more massive than this characteristic quenching scale would be quiescent.}
\label{fig:halo}
\end{figure}

Fig.~\ref{fig:halo} shows the joint evolution of  $\Sigma_1$ and halo mass for naked red nuggets where, contrary to Fig.~\ref{fig:quench}, there is a well-defined quenching scale at $\log \Sigma_1 + 0.32 \log M_{halo} \simeq 5.6$, independently of galaxy or halo mass. The coefficient of the linear combination was tuned to minimize the scatter in the $\log \Sigma_1 + \beta \log M_{halo}$ distribution at $z=0$. Since we expect the effect of mergers to dominate the evolution of massive galaxies from $z\sim2$ on, we used the SMHM at $z=2$. The characteristic $\Sigma_1$-to-$M_{halo}$ ratio is independent of this redshift choice, which only affects the coefficients of the linear combination.

Among all the parameters explored in this paper, only the combination of $\Sigma_1$ and $M_{halo}$ seems to accurately predict galaxy quenching, in agreement with previous studies \citep{Fang13,Woo15}. The build-up of the central component (i.e. $\Sigma_1$) implies an enhancement of stellar feedback in the cores of naked red nuggets, effectively affecting the energetic balance within the dark matter halos. Although this stellar feedback could in principle affect the quenching process, it becomes inefficient for halo masses above $M_\mathrm{halo} \sim 10^{12}$\msun \citep{Keller16} and hence it is not likely to be the dominant effect in our sample of massive naked red nuggets. Fig.~\ref{fig:halo} suggests therefore that it is a combination of halo and black hole or central density mass what regulates star formation in massive galaxies. This result is in agreement with independent observational results suggesting a connection between the star formation and black hole activity \citep{MN16,MN18a,Terrazas16,Terrazas17,MN18b}.

\subsection{The star formation main sequence} \label{sec:sfms}

To understand the evolution of naked red nuggets with respect to the overall population of galaxies, we can compare how our sample evolves in the SFR--M$_\star$ plane. The so-called star formation main sequence (SFMS) describes the tight relation observed between star formation and galaxy mass \citep[e.g.][]{Brinchmann04,Elbaz11,Speagle14,Schreiber15} and it is observed both locally and at high redshift \citep[e.g.][]{Noeske07,Daddi07,Wuyts11,Rodighiero11,Tacconi13}. The physical origin of the SFMS (and its scatter) is still under debate \citep[e.g.][]{Dutton10,Abramson15,Mun15,Kurczynski16,Mitra17,Matthee18,Matthee18b} but understanding its emergence, evolution and properties provides valuable insights on the baryonic cycle in our Universe. 

\begin{figure*}
\begin{center}
\includegraphics[width=14.2cm]{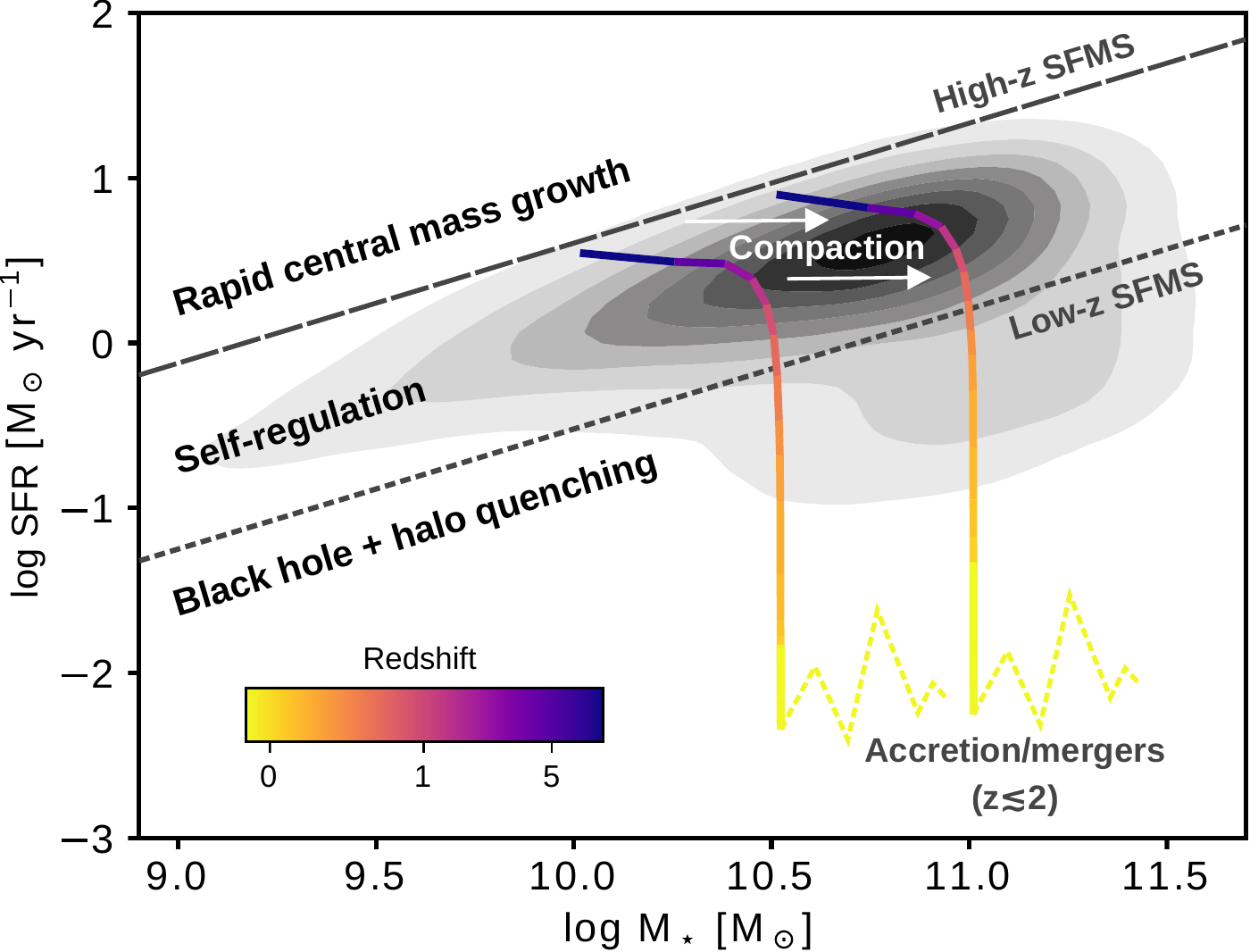}
\end{center}
\caption{Evolution of naked red nuggets across the SFMS. Contours are the $z\sim0$ measurements of \citet{Brinchmann04}, while colored curves show the measured evolution of lower-mass ($log M_\star / \msun \simeq 10.5$) and higher-mass ($log M_\star / \msun \simeq 11.0$) galaxies in our sample (color-coded by redshift). Dashed and dotted lines schematically indicate the evolutionary stages described in \S~\ref{sec:sfms}. Naked red nuggets evolve almost horizontally across the SFMS in $\sim$ Gyr scales while they are forming stars. This secular evolution would explain almost entirely the observed scatter in the SFMS, as well as the higher normalization at higher redshift (dashed lines). Following Fig.~\ref{fig:param}, the evolution across the SFMS also corresponds to a morphological compaction (i.e. higher S\'ersic index), indicated by the white arrows, which would further contribute to the quenching process. Effectively, the vertical scatter of the SFMS is determined by the time that galaxies spend forming stars, which seems to be regulated by the black hole and halo mass. Once quenched, we observe very low SFR in our sample due to the lack of recent accretion. In standard (i.e. non-relic) massive galaxies, the effect of mergers at $z\sim2$ and below would enhance the observed levels of star formation (yellow dashed lines). The three main phases of galaxy and black hole evolution described in the main text are also indicated.}
\label{fig:MS}
\end{figure*}

In Fig.~\ref{fig:MS} we show the $z\sim0$ measurements of \citet{Brinchmann04}, along with the evolutionary track followed by lower- and higher-mass naked red nuggets in our sample. A number of important features can be noticed. First, the agreement between local measurements and our inferred SFRs is remarkable. Not only we match the absolute calibration, but we also recover the observed mass dependence: for very old ages (top left points in our measured tracks) more massive galaxies are forming star more rapidly than low-mass objects. It is worth noting here that our SFR are based on the decomposition of the optical spectra of nearby galaxies into different single stellar population models. This spectroscopic-based separation is then transformed into mass ratios using the model predictions for the mass-to-light ratios, and finally averaged over the time resolution of the models (which become more coarse for older ages). Fig.~\ref{fig:MS} is an empirical test showing the strength of our approach and the quality of current state-of-the-art stellar population synthesis models. 

Second, and arguably, one of the most noticeable features in Fig.~\ref{fig:MS} is the fact that naked red nuggets follow almost horizontal (constant SFR) tracks while forming stars. A direct implication of this behavior is that up to 0.31 dex of the scatter around the SFMS could be explained by this secular process, where galaxies form stars at a roughly constant pace. The observed scatter around the SFMS would be determined by the time that it takes for a population of galaxies to reach a critical black hole-to-halo mass ratio, which quenches any further star formation. The scatter in the SFMS would be therefore reflecting differences in the long-term ($\sim$Gyr) evolutionary stages of galaxies. It is important to mention that due to the low temporal resolution for old stellar populations, we are not sensitive to short (bursty) variations in the SFR, which could also contribute to the observed scatter in the vertical direction.

An additional consequence of Fig.~\ref{fig:MS}, related to the previous point, is that the normalization of the SFMS evolves in time, which naturally explain the observed trends with redshift \citep[e.g.][]{Noeske07,Daddi07,Karim11,Rodighiero11,Whitaker12,Schreiber15,Santini17}. The oldest stellar population models in our analysis already show the SFR-M$_\star$ trend. Our results are in agreement with the recent study of \citet{Sebastian18}. Note that, particularly for low-mass naked red nuggets, these initial stages are associated with a rapid black hole growth, which would take place in the upper envelope of the SFMS \citep[e.g.][]{wellons,Pacucci15,AA17,McAlpine18}. In more massive red nuggets this growth would have happened faster and we do not have enough temporal resolution.

After being quenched, naked red nuggets in Fig.~\ref{fig:MS} show a rapid decline towards very low SFRs ($\log$\,SFR $\sim -2$), lower than the average population of galaxies at $z\sim 0$, likely due to the lack of late accretion. Although the later merger-driven evolution is thought to be mainly {\it dry} (i.e. without significant gas accretion), some fraction of (residual) star formation is expected \citep[e.g.][]{Trager00b,Vazdekis16,MN18c}, which would be particularly noticeable for very old/quiescent objects.

Finally, notice that the relation between SFR and S\'ersic index ($n$) shown in Fig.~\ref{fig:evol} and the evolutionary tracks in Fig.~\ref{fig:MS} predict that galaxies exhibit trends of constant $n$ parallel to the SFMS. This is a well-known characteristic of samples of star forming galaxies \citep[e.g.][]{Wuyts11,Brennan17}. This {\it compaction} phase, as indicated in Fig.~\ref{fig:MS}, suggests that a morphological transition occurs across the SFMS, possibly further contributing to the quenching process \citep{Martig09}.

\section{Summary and conclusions}

In this paper we have dissected the evolution of naked red nuggets. These objects, which were able to survive due to the stochastic nature of the $\Lambda$-CDM Universe, offer a unique possibility to study in great detail star formation within massive halos without dilution of information due later accretion. Using STECKMAP to measure the SFH of red nuggets from their optical spectra, we have investigated the quenching process of massive galaxies, and how it is imprinted in the internal structure of these objects. Our approach allows us to trace back the evolution of individual galaxies. This constitutes a big advantage over the standard method of analyzing slices of the Universe with increasing redshift, as we can causally connect individual measurements, getting rid of any sort of progenitor biases. The main ideas and results of this paper, summarized in Fig.~\ref{fig:MS}, are as follows:

\begin{itemize}
\item While red nuggets are star forming, their radial light distribution becomes more concentrated over time. New generation of stars form following higher S\'ersic index profiles (compaction phase in Fig.~\ref{fig:MS}), explaining the observed relation between $n$ and SFR. This initial star forming phase happens without a significant change in the effective radius. After being quenched, red nuggets keep forming stars at a very low level and in a more disky (low $n$) configuration.

\

\item Quenching seems to be determined by a combination of central stellar density and halo mass. We interpret this as a characteristic black hole-to-halo mass ratio (Fig.~\ref{fig:halo}) separating quiescent and star forming galaxies. Additionally, morphological quenching associated to high S\'ersic values might also help to stabilize gas against further fragmentation. The role of stellar feedback is not expected to be significant given the high stellar masses of naked red nuggets in our sample.

\

\item The $\Sigma_1$--M$_\star$ relation evolves in time, but only for very old stars (or high redshifts). Central density $\Sigma_1$ initially evolves faster than the stellar mass, but it then follows a track of constant $\Sigma_1$/M$_\star$. We hypothesize that this is reflecting a phase of rapid black hole growth which would happen above the main sequence. This initial stage would be then followed by a second phase where stellar and black hole mass grow in a self-regulated way, as shown in Fig.~\ref{fig:MS}.

\

\item Red nuggets follow almost horizontal tracks across the SFMS. The {\it fundamental} SFMS relation is already in place for very old stars ($\gtrsim 13$ Gyr. The horizontal tracks followed during the star-forming, self-regulated phase, would naturally induce a 0.31 dex scatter in the SFMS. The amplitude of the vertical scatter in the  SFMS is set by the time that it takes for a galaxy to reach the critical central density-to-halo mass ratio. Hence, our observations suggest that (part of) the scatter of the local SFMS is due to galaxies at different evolutionary stages. This process of constant SFR evolution also explains the higher normalization of the SFMS at higher redshifts.

\

\item Finally, our sample exhibit very low SFR at $z\sim0$ due to the absence of mergers/accretion. The evolution of normal galaxies from $z\sim2$ on would be followed by {\it dry} accretion events, which would grow their stellar masses and sizes, while slightly enhancing the SFR. This final stage, absent in naked red nuggets, is also sketched in Fig.~\ref{fig:MS}.

\end{itemize}

\section*{Acknowledgments}

We would like to thank the anonymous referee for their kind and constructive comments on the manuscript. I.M.N. acknowledges funding from the Marie Sk\l odowska-Curie Individual {\it SPanD} Fellowship 702607, and from grant AYA2016-77237-C3-1-P from the Spanish Ministry of Economy and Competitiveness (MINECO). GvdV acknowledges funding support from the European Research Council (ERC) under the European Union's Horizon 2020 research and innovation programme under grant agreement No 724857 (Consolidator Grant ArcheoDyn). I.M.N thanks David Koo, Joel Primack, Sandy Faber, Avishai Dekel, Francesco Belfiore, Cristina Ramos, Mar Mezcua, and the TRACES group for their comments and suggestions.




\bibliographystyle{mnras}
\bibliography{relics_varXiv} 


\bsp	
\label{lastpage}
\end{document}